\documentclass{webofc}
\usepackage[varg]{txfonts}   % Web of Conferences font
\usepackage{hyperref}
\usepackage{url}
\hypersetup{colorlinks=true,citecolor=blue,urlcolor=blue,linkcolor=blue}
\begin{document}

\null\hfill\begin{tabular}[t]{l@{}}
  \text{MIT-CTP/5870}
\end{tabular}
\title{Holographic Heavy Quark Energy Loss in the Hybrid Model}

\author{\firstname{Jean F.} \lastname{Du Plessis}\inst{1}\fnsep\thanks{\email{jeandp@mit.edu;} speaker at Hard Probes 2024} \and
        \firstname{Daniel} \lastname{Pablos}\inst{2}\fnsep\thanks{\email{daniel.pablos@usc.es}} \and
        \firstname{Krishna} \lastname{Rajagopal}\inst{1}\fnsep\thanks{\email{krishna@mit.edu}}
}

\institute{Center for Theoretical Physics, Massachusetts Institute of Technology, Cambridge, MA 02139 
\and
IGFAE, Universidade de Santiago de Compostela, E-15782 Galicia-Spain
%          Instituto Galego de F\'ısica de Altas Enerx\'ıas IGFAE, Universidade de Santiago de Compostela,
%E-15782 Galicia-Spain
          }

\abstract{To date, holographic calculations in strongly coupled plasma have provided separate descriptions for the rates of energy loss either for ultrarelativistic massless quarks and gluons or for infinitely massive quarks, with the latter calculation valid for $\sqrt{\gamma} < M/(\sqrt{\lambda}T)$, where $\gamma$ is the Lorentz boost factor for a heavy quark with velocity $v$ and mass $M$ moving through plasma with ’t Hooft coupling $\lambda$ and temperature $T$. These two calculations should apply sequentially in the description of the energy loss of a heavy quark that starts out ultrarelativistic, loses energy, slows down, becomes non-relativistic at later times, and ultimately comes to rest and diffuses in the strongly coupled plasma. We provide an ansatz for uniquely incorporating both regimes to give an approximate but unified description of how a heavy quark that is initially ultrarelativistic loses energy all the way until it comes to rest. We implement this ansatz in the Hybrid Strong/Weak Coupling Model. With this new, consistent, treatment of heavy quark energy loss at strong coupling, we confront our predictions for the suppression and azimuthal anisotropies of D- and B-mesons, as well as B-tagged jets, with available experimental data.}
\maketitle
\section{Introduction}
\label{intro}
With the strongly coupled nature of Quark-Gluon Plasma (QGP) at the temperatures reached in heavy ion collisions well established, we wish to model how heavy quarks lose energy in strongly coupled plasma. At strong coupling, perturbative techniques break down, leaving us in need of alternative approaches. For problems that can be formulated in Euclidean time one can use lattice QCD techniques. An example of this is calculations of the spatial diffusion coefficient, which can be related to the drag force experienced by a heavy quark in the $v\to0$ limit. For dynamic processes such as energy loss at nonzero velocity, however, lattice QCD calculations in general do not provide insight. For such questions, we can turn to the AdS/CFT duality in order to calculate quantities of interest in the strongly coupled plasma of ${\mathcal N}=4$ SYM theory. This allows us to determine the parametric dependence of heavy quark energy loss from first principles in a related but different strongly coupled gauge theory at nonzero temperature. In this work we shall employ existing AdS/CFT calculations of the energy loss of light quarks where the heavy quark is sufficiently relativistic as well as of the late-time behavior of  (infinitely) heavy quarks. We will implement the resulting composite description of heavy quark energy loss in the hybrid strong/weak coupling model~\cite{Casalderrey-Solana:2014bpa,
%Casalderrey-Solana:2015vaa,
Casalderrey-Solana:2016jvj,
Hulcher:2017cpt,
Casalderrey-Solana:2018wrw,
Casalderrey-Solana:2019ubu}.
%,Hulcher:2022kmn}.

\section{Composite description of heavy quark energy loss}
In strongly coupled $\mathcal{N}=4$ SYM theory, there are holographic calculations of the energy loss of fundamental flavors in two regimes: massless, and infinitely massive.
One can numerically calculate finite mass corrections to the infinite mass calculation, but one still assumes that the quark is being pulled at constant velocity and has been losing energy forever.

In principle it is possible to do a holographic calculation of a finite mass quark losing energy to the plasma, but this has not been done. For phenomenological purposes, however, we can employ the results obtained in the above two limits to obtain a reasonable composite description of a heavy quark that is initially relativistic, loses energy, and comes to rest.

In the massless limit, one finds the following formula for 
energy loss~\cite{Chesler:2014jva,Chesler:2015nqz} :
\begin{equation}\label{eq:lightenergyloss}
  \frac{d E}{dx}=-\frac{4 x^2 E_0}{\pi x_t^2\sqrt{x_t^2-x^2}},
\end{equation}
and in the infinite mass limit, one finds \cite{Herzog:2006gh,Gubser:2006bz}  
\begin{equation}\label{eq:drag}
  \frac{dp}{dt}=-\eta_D p.
\end{equation}
Here $E_0$ is the initial energy, $x$ is the length of plasma traversed,
\begin{equation}
  x_t\equiv \frac{1}{2\kappa_{sc}}\left(\frac{E_0}{T^4}\right)^\frac{1}{3}
\end{equation}
is the thermal stopping length of a light quark, $t$ is time elapsed in the fluid rest frame, and
\begin{equation}
  \eta_D\equiv\kappa_{HQ}\frac{T^2}{M}
\end{equation}
is the heavy quark drag coefficient and $M$ is the mass of the heavy quark. Here, $\kappa_{sc}$ and $\kappa_{HQ}$ are parameters that govern the strength of the coupling between the light and heavy quarks with the strongly coupled plasma. Both can be calculated in ${\mathcal N}=4$ SYM theory.  For our purposes, in modeling heavy quark energy loss in the QCD QGP produced in heavy ion collisions both are free parameters that can be fit to data (as $\kappa_{sc}$ has been~\cite{Casalderrey-Solana:2018wrw}) or 
varied to estimate theoretical uncertainty.
We can convert the drag equation \eqref{eq:drag} into an energy loss formula using $E^2=p^2+M^2$ and $\frac{d}{dt}=\frac{dx}{dt}\frac{d}{dx}=\frac{p}{E}\frac{d}{dx}.$
This gives
\begin{equation}\label{eq:heavyenergyloss}
  \frac{d E}{d x}=-\eta_D\sqrt{E^2-M^2}\ ,
\end{equation}
which can be employed to describe a heavy quark which is not being pulled and which therefore loses energy and comes to rest~\cite{Herzog:2006gh}.

It is easy to verify from Eqs.~\eqref{eq:lightenergyloss} and \eqref{eq:heavyenergyloss} that the second derivative of the energy of a light quark is always negative, while that of a heavy quark is always positive.
This means that for a given $E_0$ there is at most one point where the two energy loss formulas agree. In general we find that there is always a point where this happens, since the light quark energy loss starts at 0 and diverges to $-\infty$ in finite length $x$.
This leads us to propose the following composite model for phenomenological purposes.
For a heavy quark that starts off energetically, one would expect it to lose energy like a light quark early on, and as a late time heavy quark at asymptotically late times,
so we can simply take 
\begin{equation}\label{eq:composite}
  \frac{dE}{dx}=-\min\left(\frac{4 x^2 E_0}{\pi x_t^2\sqrt{x_t^2-x^2}},\,\eta_D\sqrt{E^2-M^2}\right),
\end{equation}
which will give a continuous and  once-differentiable $E(x)$ function by construction; note that it will have a discontinuous second derivative at the point where it switches from light to heavy quark energy loss.
This composite description should capture the early- and late-time behavior of a heavy quark well. We will implement Eq.~\eqref{eq:composite} in the Hybrid 
Model~\cite{Casalderrey-Solana:2014bpa,
%Casalderrey-Solana:2015vaa,
Casalderrey-Solana:2016jvj,
Hulcher:2017cpt,
Casalderrey-Solana:2018wrw,
Casalderrey-Solana:2019ubu},
%,Hulcher:2022kmn}
allowing us to implement heavy quark energy loss in this model for the first time and study its observable consequences.
% One can see this as the quark `choosing' whether it wants to lose energy as a light or heavy quark in such a way that it always loses the least amount of energy possible \DP{not sure about how this sounds...}. 
% We will call this anzatz `composite energy loss'.
% , inspired by the stitching together of two seemingly incompatible pieces, as well as the shape of the solutions of the differential equation.

\section{FONLL Reweighting}
\begin{figure}
    \centering
    \includegraphics[width=0.45\linewidth]{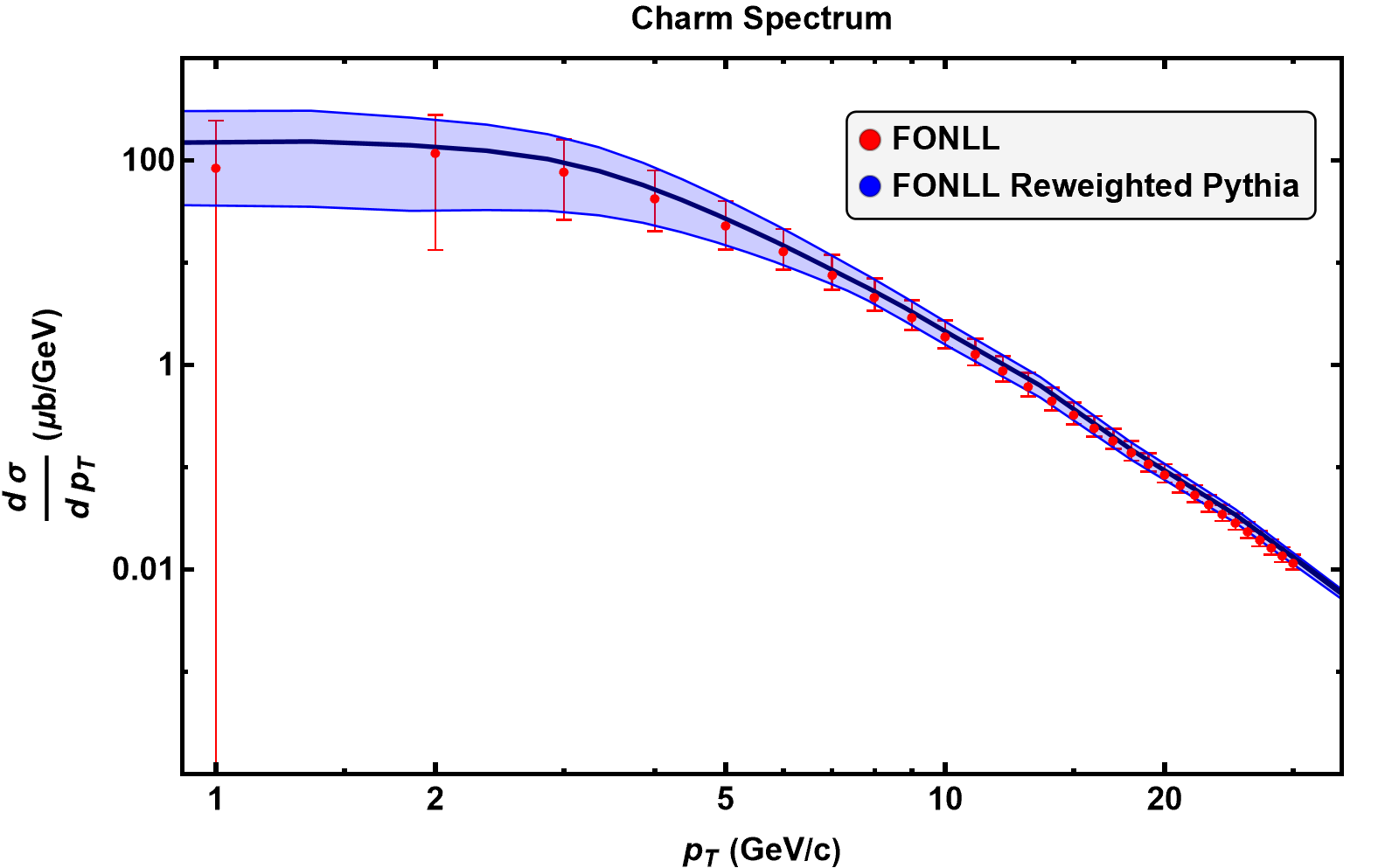}
    \includegraphics[width=0.45\linewidth]{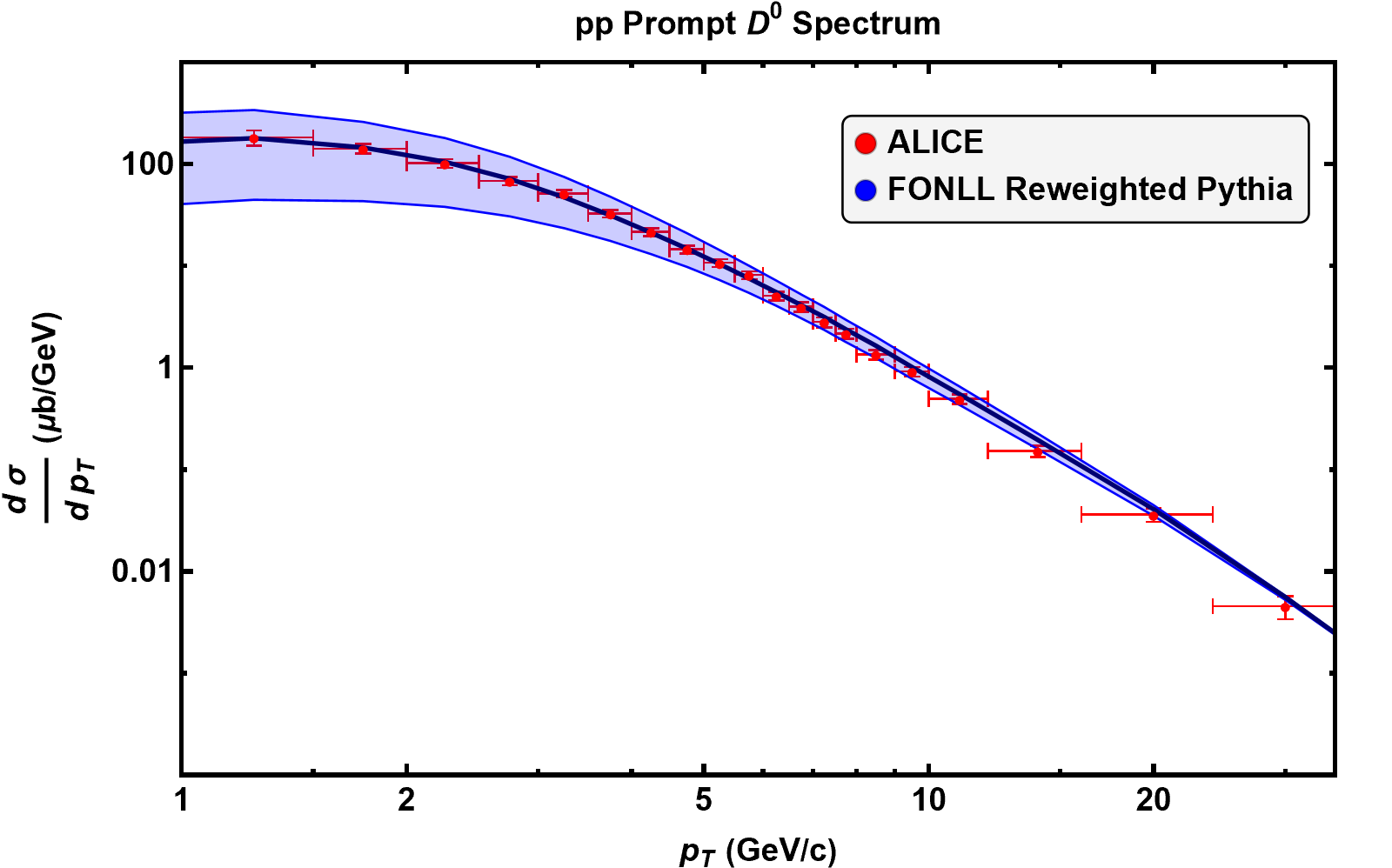}
    \caption{(Left) Charm spectrum as predicted by FONLL calculation compared to our reweighted PYTHIA8. The reweighting is such that the error bands should approximate the error bars of the FONLL calculation. (Right) Prompt $D^0$ spectrum of reweighted PYTHIA8 compared to ALICE data~\cite{ALICE:2021mgk}. Error band quantifies FONLL uncertainty; statistical uncertainty of central value is less than the thickness of the dark blue line.}
    \label{plt:fonll}
\end{figure}

At low $p_T$ the PYTHIA8~\cite{Bierlich:2022pfr} charm quark spectrum is known to fail to describe pp observables involving charmed hadrons.
%without taking FONLL calculations into account. 
One typically remedies this by reweighting events as a function of the highest $p_T$ charm quark  in the event using FONLL calculations. The FONLL spectrum \cite{Cacciari:1998it} has uncertainties coming from scale and NPDF uncertainties, so we will need slightly different reweighting schemes to represent the top of the error bars, the central value and the bottom of the error bars. We will then present all observables with a darker error band representing the statistical error after reweighting by the central value, as well as a lighter but larger error band signifying the uncertainty in the FONLL reweighting procedure, obtained from an envelope of the statistical error bars of the top and bottom reweighted results. Note that for $R_{\rm AA}$, for example, the FONLL uncertainty is correlated, so the ratios of the pairs of spectra with different reweighting is taken first, and then the envelope is calculated after. We can now compare the (reweighted) PYTHIA8 charm spectrum to that obtained directly from FONLL predictions, as well as compare the spectra of prompt $D^0$ mesons obtained in vacuum from PYTHIA8 reweighted by FONLL to that measured by ALICE in LHC heavy ion collisions~\cite{ALICE:2021mgk} --- see Fig.~\ref{plt:fonll}.
Given the excellent agreement between the central reweighted value and the experimental $D^0$ spectrum in pp however, we will take the FONLL uncertainties (indicated with lighter bands throughout) to be of auxiliary importance.

% Due to our approach of reweighting as a function of the largest charm $p_T$, we have more limited control over reweighting very low $p_T$ ($<2$GeV) charms, who will mostly be in events with a charm of higher $p_T$. We also note that we use PYTHIA8 with $\hat{p}_{t\,\text{min}}=2$GeV, further restricting the accuracy of our spectra in this very low $p_T$ region. This is expected to have a relatively limited effect, since for $R_{AA}$ ratios of spectra will mean that inaccuracies in the individual spectra cancel, and for $v_2$ many of our outgoing low $p_T$ charms will have been produced at higher $p_T$ where the spectrum is accurate. \DP{I would say that given that our pp spectrum is very well reproduced down to 1 GeV, we can forget about all these caveats?} 
% \JF{
%Given the excellent agreement between the central reweighted value and the experimental $D^0$ spectrum in pp however, we will take the FONLL uncertainties (indicated with lighter bands throughout) to be of auxiliary importance.
% } {\color{red}{Krishna suggests removing this paragraph entirely, with the exception of the last sentence that is currently in blue -- which should be kept and can be placed as the last sentence of the paragraph above.}}

\section{Results}

\subsection{Jet $R_{\rm AA}$}
\begin{figure}
    \centering
    \includegraphics[width=0.85\linewidth]{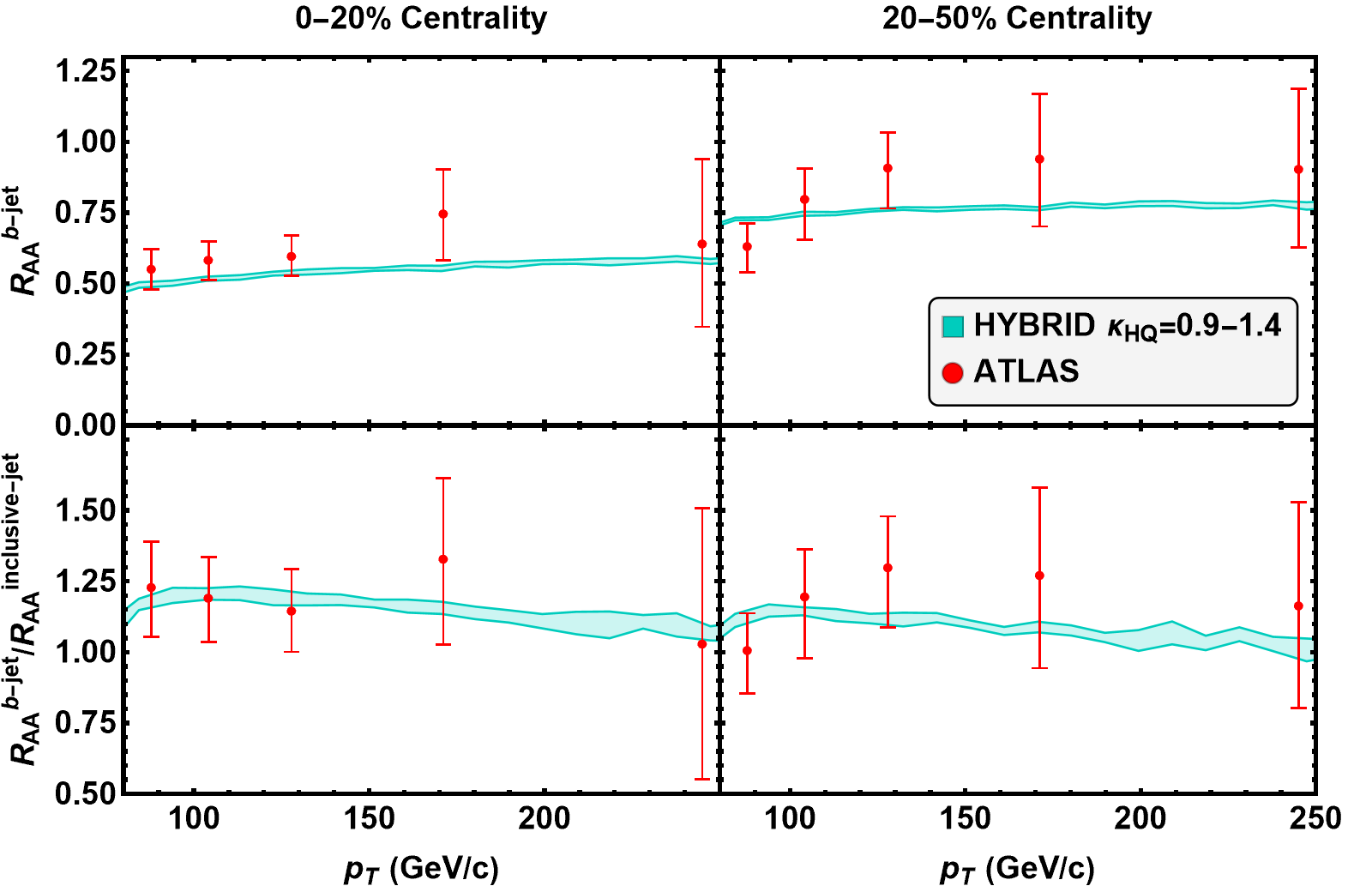}
    \caption{Comparing Hybrid Model results for $b$-tagged jet $R_{\rm AA}$ as well as for the double-ratio between $R_{\rm AA}$ for $b$-jets and for inclusive jets to ATLAS data \cite{ATLAS:2022agz}. 
    Error bands on the calculations are statistical; we have checked that variation of $\kappa_{HQ}$ over the range from 1.7 to 4.4 has no visible effect on the results.
    %Width of the error bands include both statistical uncertainty as well as the variation of the heavy-quark energy loss parameter $\kappa_{HQ}=0.9-1.4$.
    }\label{plt:bjet}
\end{figure}

By construction, high-$p_T$ heavy quarks behave indistinguishably from light quarks in our newly presented composite description of energy loss. 
It is such energetic heavy quarks that enter the calculation of the $b$-tagged jet $R_{\rm AA}$ observable, which we present in Fig.~\ref{plt:bjet}. We have checked that the results are completely insensitive to the value of $\kappa_{HQ}$ chosen, which confirms the intuition that this observable does not probe the heavy quark energy loss part of our model. We also note that the good agreement with ATLAS data \cite{ATLAS:2022agz} of both the $b$-jet $R_{\rm AA}$ and the $R_{\rm AA}$ double-ratio of that in $b$-jets to that in inclusive jets is a test of the handling of the strongly coupled energy loss of quarks vs gluons in the Hybrid Model.

\subsection{Hadron $R_{\rm AA}$ and $v_2$}

\begin{figure}[t]
    \centering
    \includegraphics[width=0.45\linewidth]{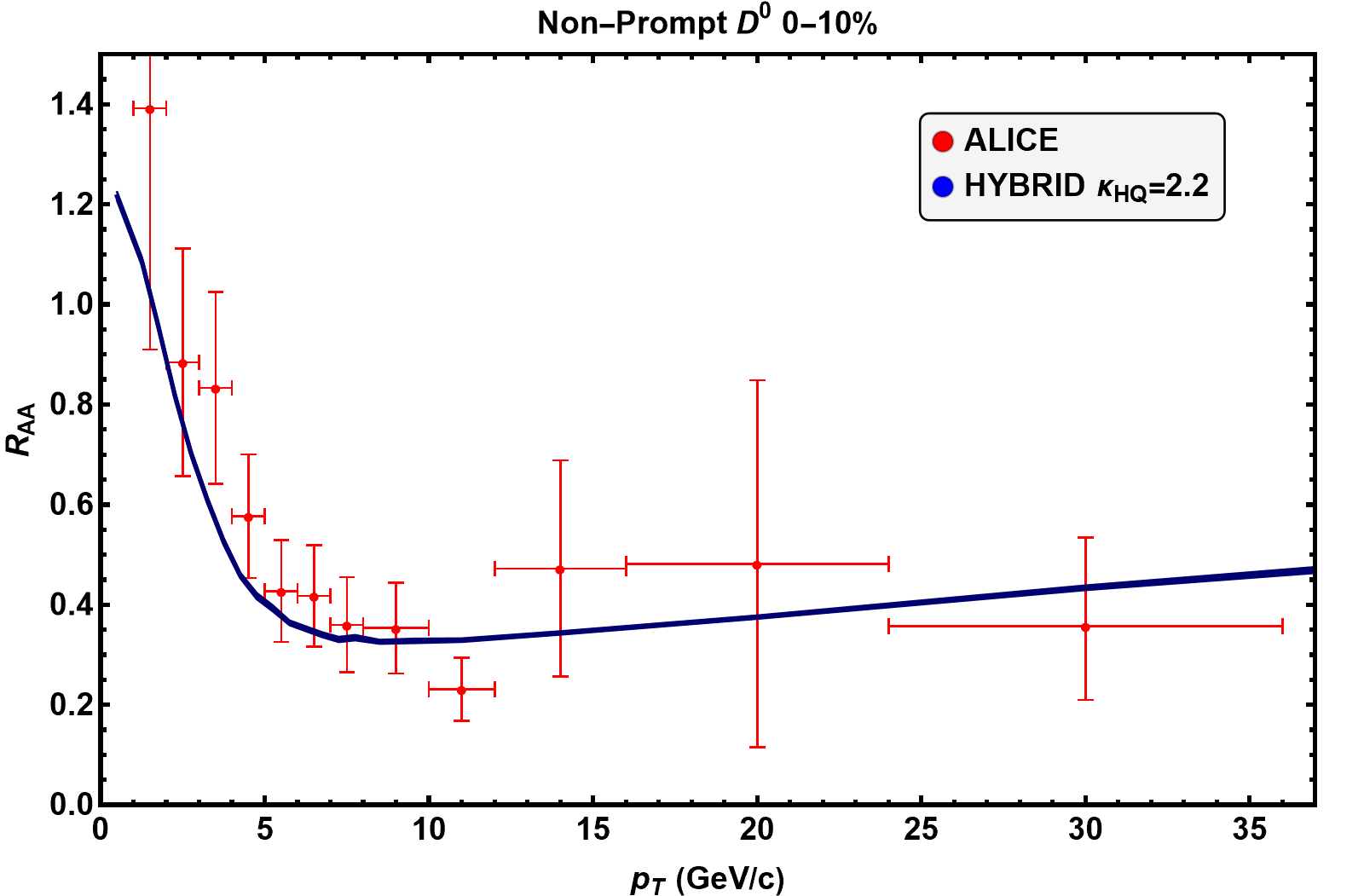}
    \includegraphics[width=0.45\linewidth]{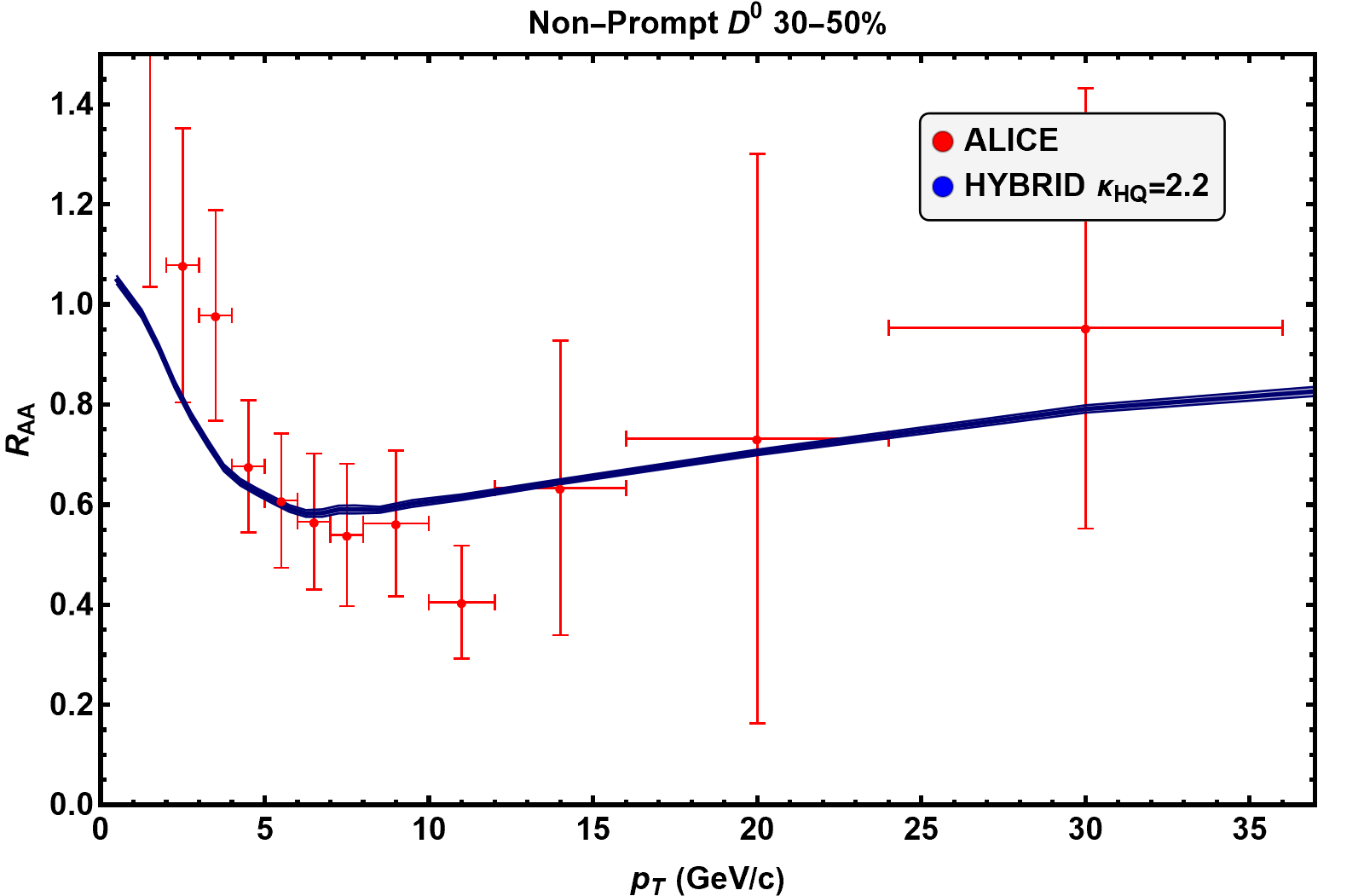}\\
    \includegraphics[width=0.45\linewidth]{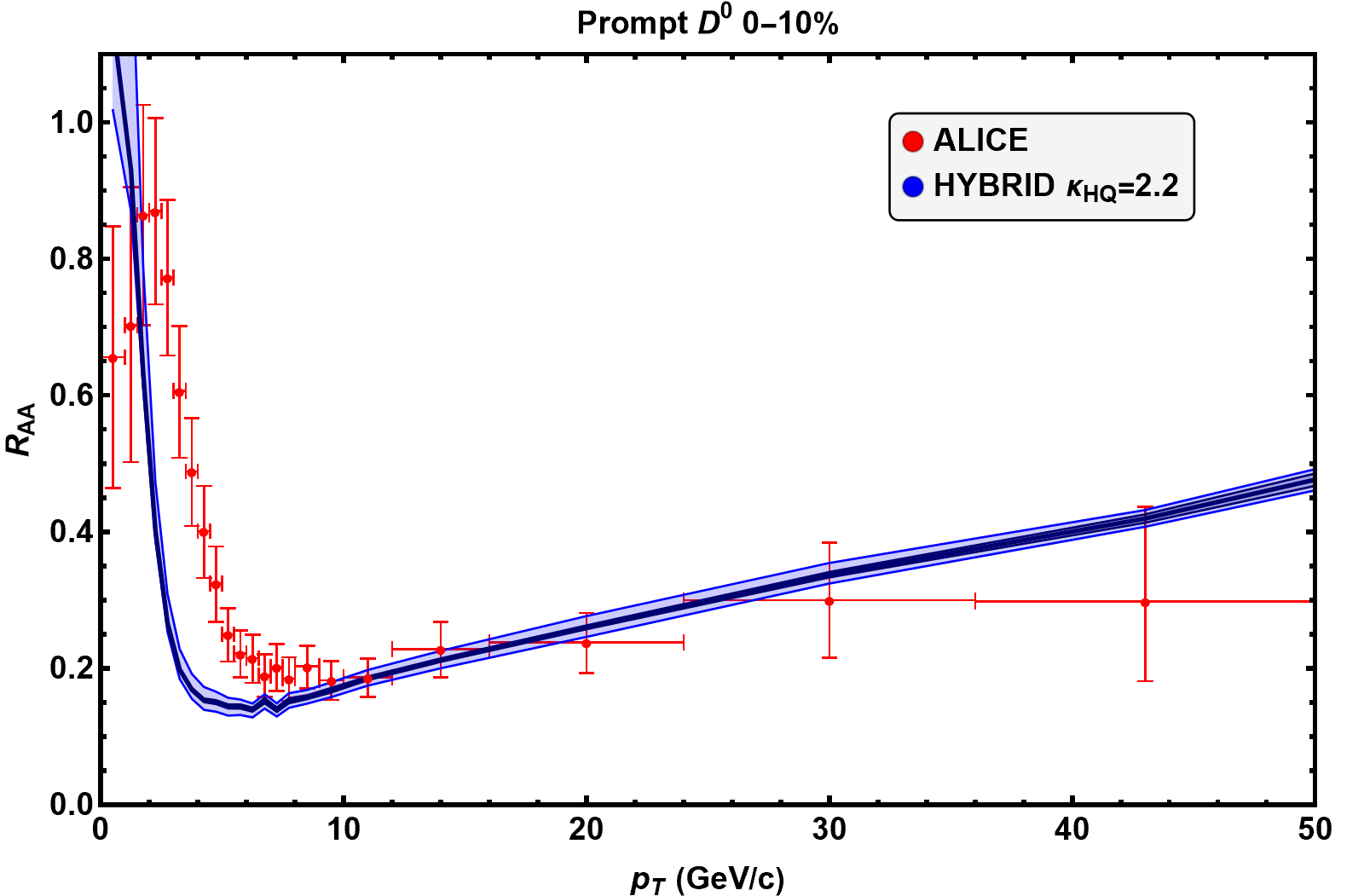}
    \includegraphics[width=0.45\linewidth]{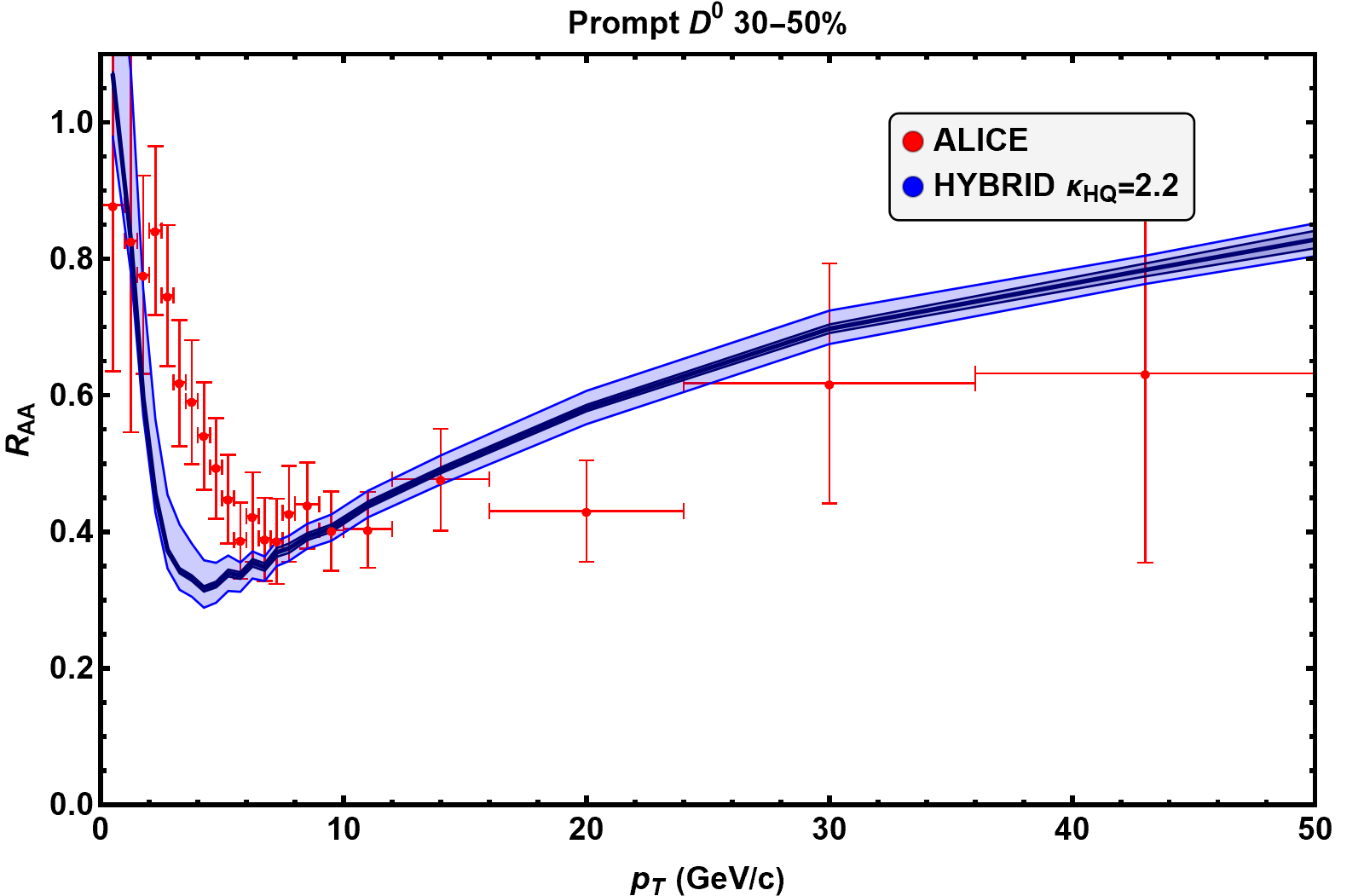}
    \caption{Hybrid Model results with $\kappa_{HQ}=2.2$ for $B$-meson (non-prompt $D^0$'s; upper panels) and $D$-meson (prompt $D^0$'s; lower panels) $R_{\rm AA}$ compared to ALICE data~\cite{ALICE:2021rxa} for 0-10\% (left panels) and 30-50\% (right panels) centrality LHC collisions with $\sqrt{s_{\rm NN}}=5.02$~TeV.}
    \label{fig:hadraa}
\end{figure}

\begin{figure}
    \centering
    \includegraphics[width=0.45\linewidth]{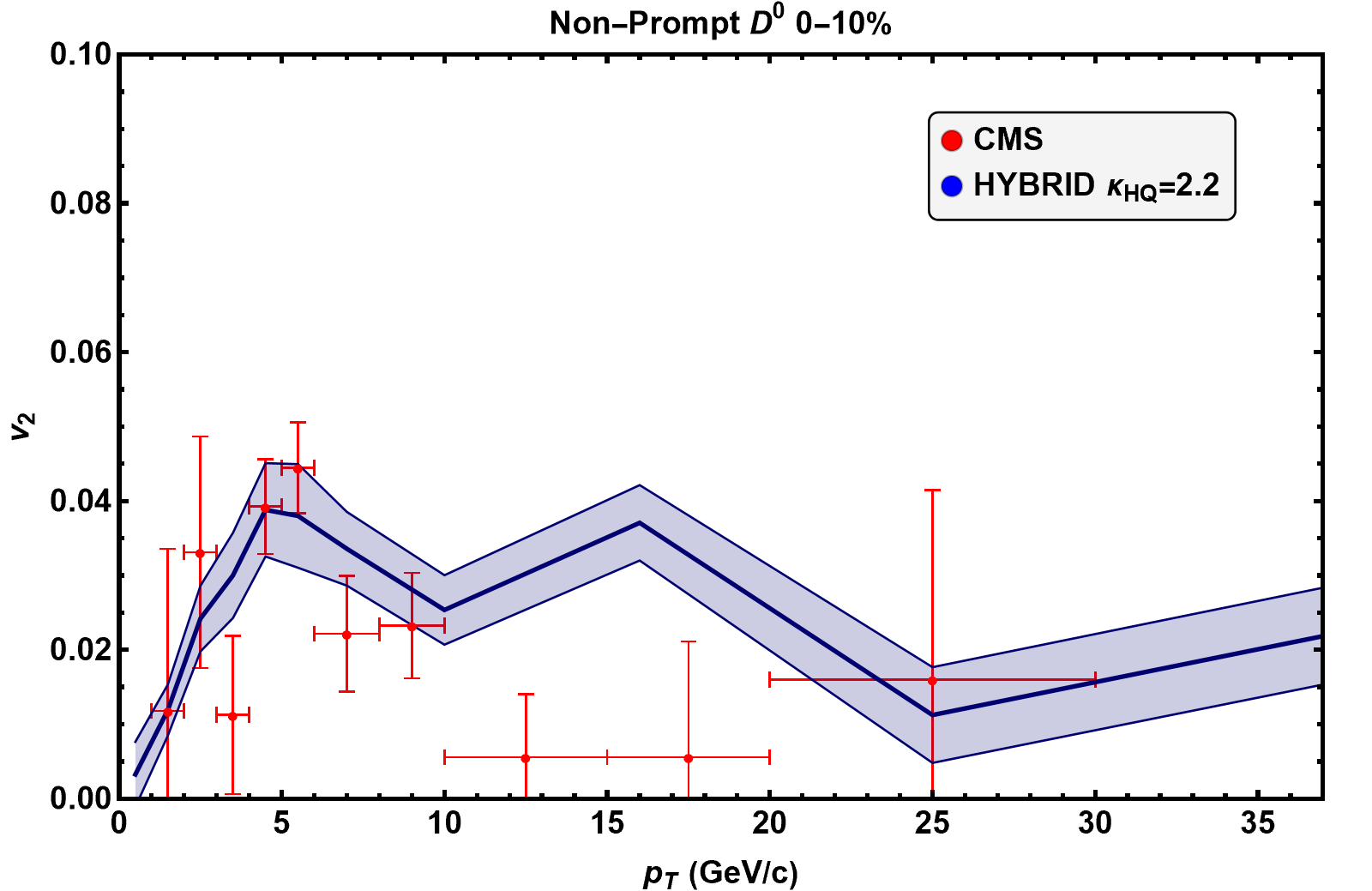}
    \includegraphics[width=0.45\linewidth]{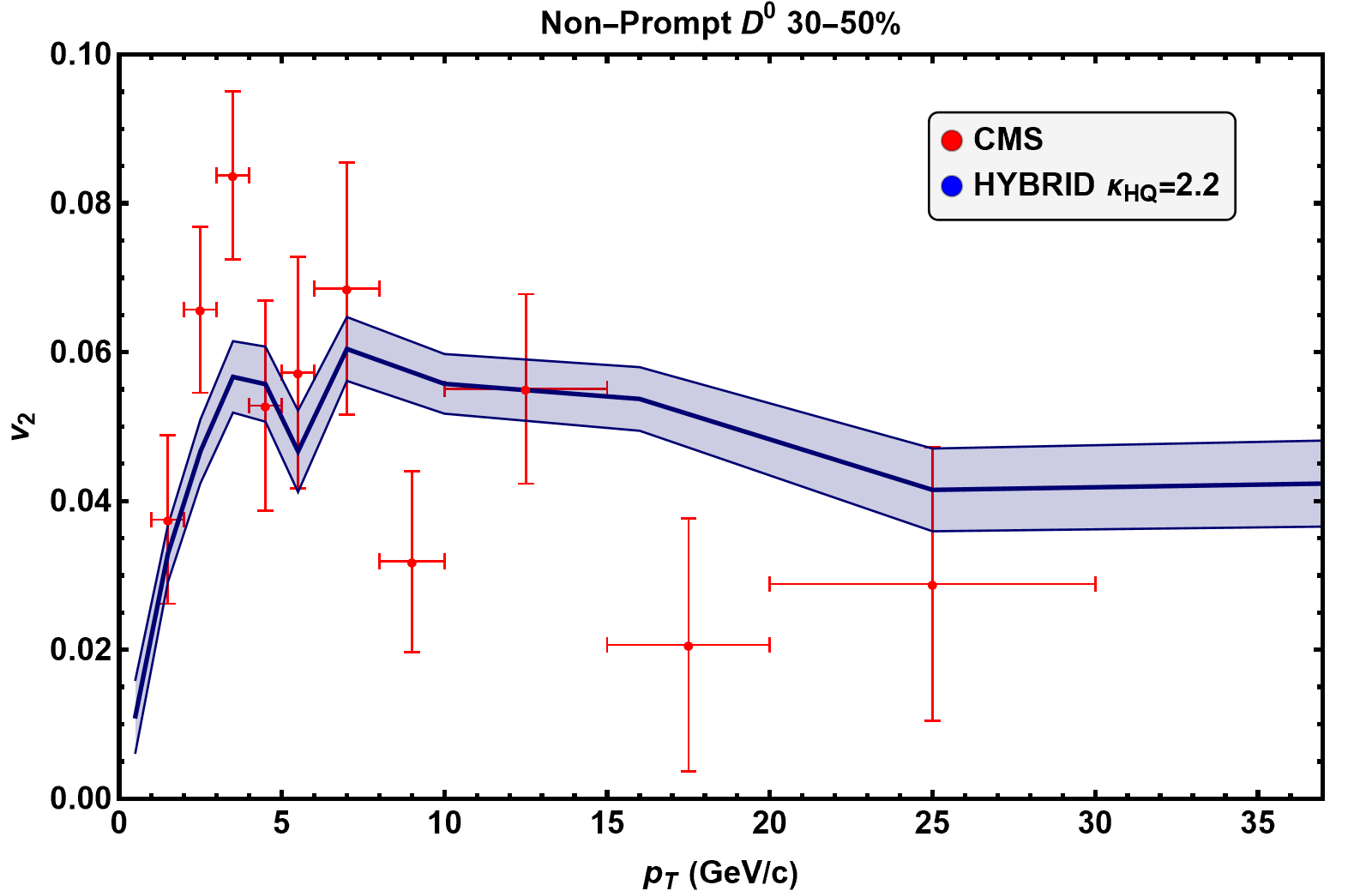}\\
    \includegraphics[width=0.45\linewidth]{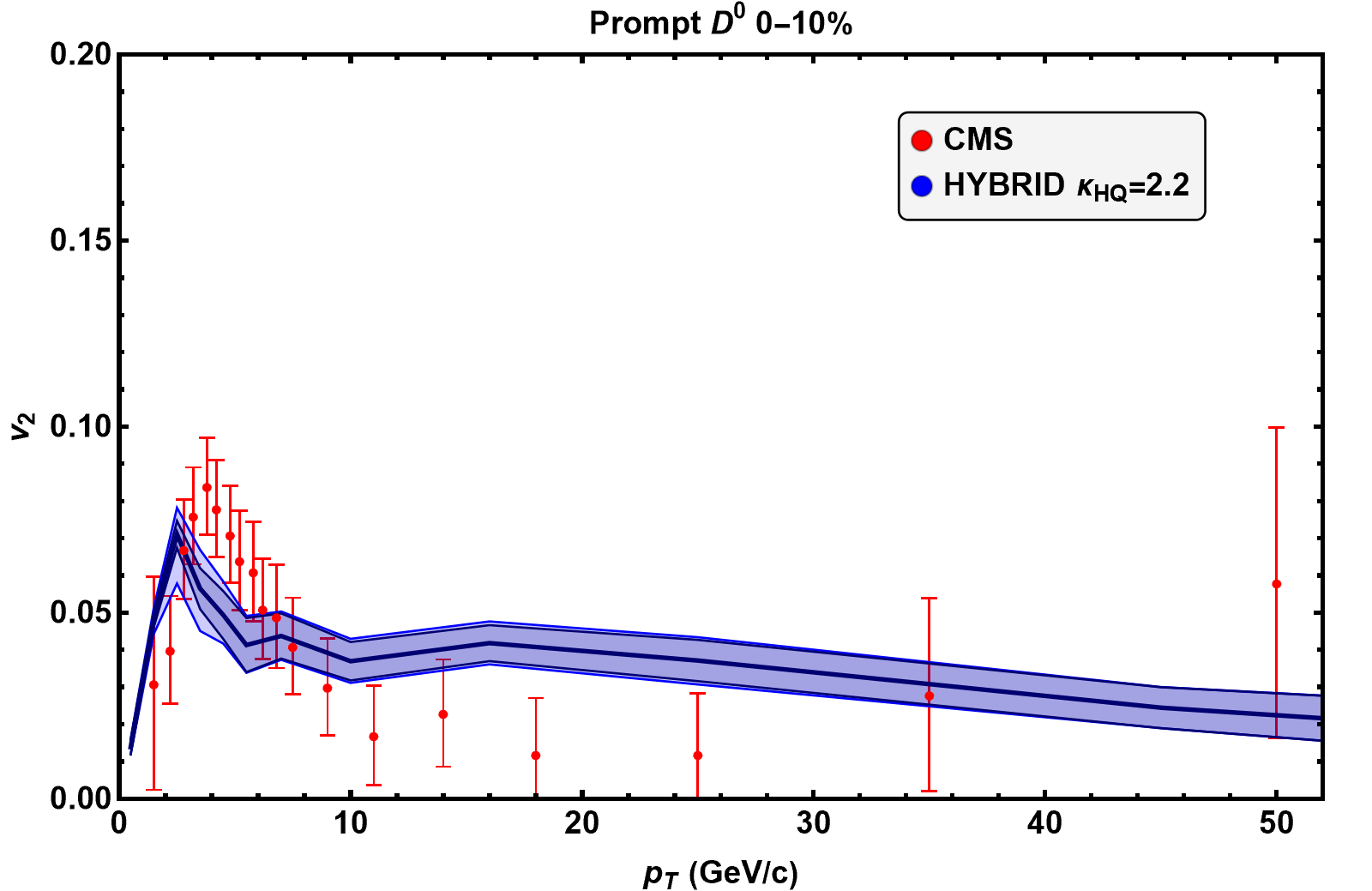}
    \includegraphics[width=0.45\linewidth]{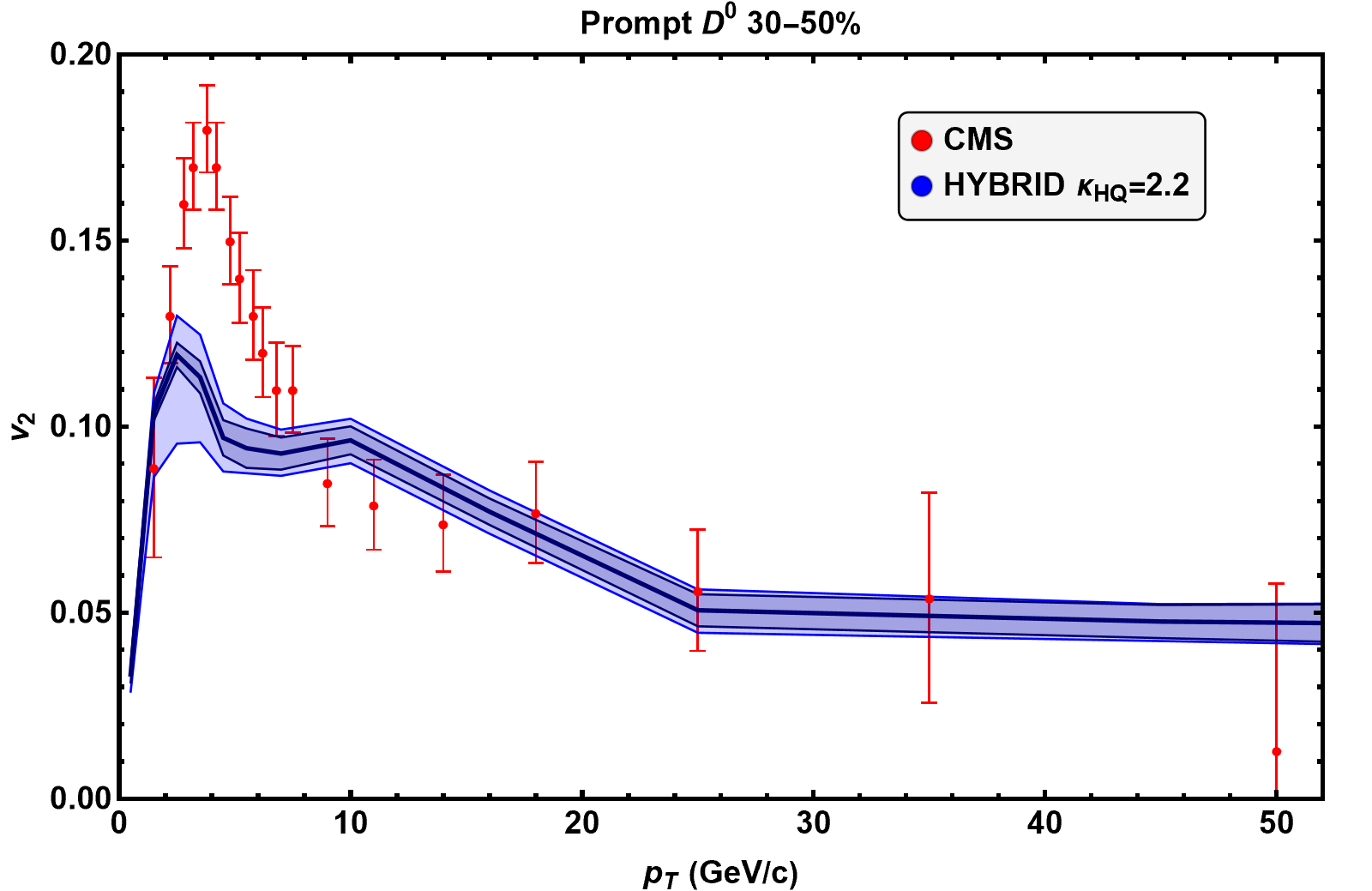}
    \caption{Hybrid Model results with $\kappa_{HQ}=2.2$ for $B$-meson (non-prompt $D^0$'s; upper panels) and $D$-meson (prompt $D^0$'s; lower panels) $v_2$ compared to CMS data~\cite{CMS:2020bnz} for 0-10\% (left panels) and 30-50\% (right panels) centrality LHC collisions with $\sqrt{s_{\rm NN}}=5.02$~TeV.}
    \label{fig:hadv2}
\end{figure}

We compare our predictions of the $R_{\rm AA}$ and $v_2$ for non-prompt and prompt $D^0$ mesons (in effect, $B$-mesons and $D$-mesons) to ALICE data~\cite{ALICE:2021rxa} in Figs.~\ref{fig:hadraa} and \ref{fig:hadv2}, respectively. 
In both Figures, we show results for a value of $\kappa_{HQ}=2.2$ which was chosen in order to be simultaneously reasonably consistent with $R_{AA}$ and $v_2$ for central collisions. This is by no means a statistical fit, something we leave to future work. 
Before attempting such a fit, we shall first need to improve our description of the $B$- and $D$-meson spectra and $v_2$ at low $p_T$, as this is where the sensitivity to $\kappa_{HQ}$ lies and this is also where at present our model fails to describe the $D$-meson data. 
We do find reasonable agreement between the predictions of our model and the experimental data 
on both $R_{\rm AA}$ and $v_2$ at large $p_T$, where heavy quarks lose energy in the same way that light quarks do. At low $p_T$ we find reasonable agreement with the $B$-meson measurements, but not for the $D$-mesons. 

%For B meson $v_2$ we find that despite the observed agreement the experimental statistical uncertainties are too large to make definitive statements. 

%For C mesons we again find reasonable agreement at high $p_T$, but at low $p_T$ we can clearly see that we are missing the expected flow effects coming from the recombination of the charm quark with a quark from the flowing medium. 

The disagreement  between the predictions of the present model for low-$p_T$ $D$-mesons and experimental data is not unexpected.
Our model as currently constituted describes the hadronization of $D$ or $B$ mesons as if the only way that they can form is if a $c$ or $b$ quark in a Hybrid Model parton shower hadronizes with a light antiquark from the same parton shower. In reality, the heavy quarks may hadronize this way but it is also possible for them to hadronize with a light quark from the QGP, a process referred to as recombination or coalescence.
Because the hydrodynamic droplet of QGP is expanding radially outwards at the time of freezeout, recombination transfers momentum from the flowing medium to the $D$- or $B$-meson. This will tend to increase both $R_{\rm AA}$ and $v_2$ at low-$p_T$, with the effects being smaller for $B$-mesons because they are more massive.
%expected given the present absence of recombination effects in our model, which are known to specially affect $D$ mesons at low $p_T$. 
Such effects will be included in a forthcoming publication.

%We interpret this as signaling the need for recombination effects at these low transverse momenta for the lighter charm quark which is more significantly affected by it.

\section{Conclusions}

We have formulated a composite description of the energy loss of heavy quarks in strongly coupled plasma using existing AdS/CFT calculations in the massless and infinite mass limits, allowing us to include heavy quarks in the Hybrid Model for the first time. High-$p_T$ observables such as b-jet $R_{AA}$ are only sensitive to the regime in which the $b$-quarks lose energy as light quarks do, and yield reasonable results when confronted with experimental data. $D$-meson $R_{\rm AA}$ and $v_2$ observables 
agree well with data all the way down to $p_T\gtrsim 10$~GeV, for the same reason.
%at lower $p_T$ ($p_T > 10 \text{GeV}$) agree well with data, but is still largely insensitive to the heavy quarks mass. 
We see much larger sensitivity to the regime in which the $c$-quarks lose energy as heavy quarks do at lower $p_T$, but also observe that our model neglects physical effects such as recombination which are relevant in this regime. This limits our ability to use the confrontation between Hybrid Model calculations and experimental measurements of $D$-meson spectra and flow to constrain the value of $\kappa_{\rm HQ}$, the parameter which governs heavy quark energy loss, until such effects are included. We plan to do so in future work.\\

%. which limits the interoperability of our results in this region until such effects are included in future work.

We are grateful to Carlos Hoyos and Bruno Scheihing-Hitschfeld for helpful conversations.
Research supported in part by the U.S.~Department of Energy, Office of Science, Office of Nuclear Physics under grant Contract Number DE-SC0011090.
Research of DP supported in part by the European Union's Horizon 2020 research and innovation program under the Marie Sk\l odowska-Curie grant agreement No 101155036 (AntScat), by the European Research Council project ERC-2018-ADG-835105 YoctoLHC, by the Spanish Research State Agency under project 
PID2020-119632GB-I00, by Xunta de Galicia (CIGUS Network of Research Centres) and the European Union, and by Unidad de Excelencia Mar\'ia de Maetzu under project CEX2023-001318-M.

\bibliography{qgp_EEC_ref.bib}

\end{document}